\documentclass[twocolumn,showpacs,preprintnumbers,amsmath,amssymb]{revtex4}

\usepackage{makeidx}
\usepackage{verbatim,vmargin,amsmath,calc, mathrsfs}
\usepackage{amsthm, amssymb, dsfont}
\usepackage{epsfig, pstricks, pst-plot,pst-node}
\usepackage{graphicx}
\usepackage{dcolumn}
\usepackage{bm}

\begin{document}
\newcommand{\yps}{YbPd$_2$Sn\ }
\newcommand{\simi}{$\sim$}

\preprint{APS/123-QED}

\title{Pressure induced changes in the \\antiferromagnetic superconductor \yps}

\author{A-M. Cumberlidge}
\email{amc71@cam.ac.uk}
\author{P.L. Alireza}%
\author{A.F. Kusmartseva}
\author{E.E. Pearson}
\author{G.G. Lonzarich}
\affiliation{%
Cavendish Laboratory, Cambridge University, J. J. Thomspn Av., Cambridge CB3 0HE, UK
}%

\author{N.R. Bernhoeft}
\affiliation{18 Maynestone Road, Chinley, SK23 6AQ, UK}
\author{B. Roessli}
\affiliation{SINQ @ PSI Villigen, Suisse}

%

\date{\today}

\begin{abstract}
Low temperature ac magnetic susceptibility measurements of the
coexistent antiferromagnetic superconductor \yps\ have been made in hydrostatic 
pressures $\leq$  74 kbar in moissanite anvil cells.  The
superconducting transition temperature is forced to T$_{SC}$ = 0 K at a
pressure \simi\ 58 kbar.  The initial suppression of the superconducting transition temperature 
is corroborated by lower hydrostatic pressure (p $\leq$ 16 kbar)
four point resisitivity measurements, made in a piston cylinder pressure cell. 
 At ambient pressure, in a modest magnetic field  of \simi\ 500 G, 
this compound displays reentrant superconducting behaviour.  
This reentrant superconductivity is suppressed to lower 
temperature and lower magnetic field as pressure is increased.  
The antiferromagnetic ordering temperature,
which was measured at T$_N$ = 0.12 K at ambient pressure is enhanced, to
reach T$_N$ = 0.58 K at p = 74 kbar. The reasons for the coexistence 
of superconductivity and antiferromagnetism is discussed in the light of 
these and previous findings.  Also considered is why superconductivity on
 the border of long range magnetic order is so much rarer in  Yb
 compounds than in Ce compounds. The presence of a new transition
visible by ac magnetic susceptibility under pressure and in magnetic
fields greater than 1.5 kG is suggested. 
\end{abstract}

\maketitle

\section{\label{Introduction}Introduction}

Heavy fermion superconductivity 
has now been observed in over twenty 
cerium based compounds (for recent reviews see Refs. \cite{Thalmeier,  Onuki, Thompson, Mathur}).
  All of these materials are examples of 
superconductors on the border of antiferromagnetic order.  
In the Ce based heavy fermion
superconductors, the Ce has nominally one f-electron (i.e. 4f$^1$).  
Corresponding ytterbium heavy fermion systems have around one electron missing from
their f-shell, thereby being the hole equivalent of the Ce in the
heavy fermion compounds. One may therefore expect to find heavy 
fermion superconductors based on Yb in the 4f$^{13}$
electron configuration.  However, such behaviour in Yb (4f$^{13}$)
systems actually proves to be exceedingly rare. As such, the search
for heavy fermion Yb based superconductors in the presence of strong
magnetic fluctuations continues.

One rare example of an Yb based superconductor with coexistent long-range
 antiferromagnetic order is \yps \cite{Kierstead}. In the case of the Ce 
heavy fermion compounds (with 4f$^1$),
strong hybridisation between the f-electrons and
 conduction electrons is a key ingredient to the occurrence of
unconventional superconductivity.  Here, the electrons responsible
for magnetism are the same ones which cause superconductivity. 
However, at first sight the
behaviour of \yps appears like it could be more reminiscent of other compounds, i.e. 
the rare earth ternary compounds, including the Chevrel phase
\textbf{M}Mo$_{6}$S$_{8}$ (\textbf{M} = Gd, Tb, Dy, and Er), some
borocarbides (\textbf{R}NiB$_{2}$C with \textbf{R} = Tm, Er, Ho and
Dy) and the rhodium boride  SmRh$_{4}$B$_{4}$ \cite{Machida1}.  
Here, the electrons
on different sublattices are responsible for the two types of
behaviour. \yps is additionally in the
same class of materials as ErPd$_{2}$Sn \cite{Stanley}.  It is a 
Heusler compound which has the cubic Cu$_{2}$MnAl-type structure
\cite{Aoki}. It contains a significantly larger percentage of
magnetic rare earth ions than other magnetic superconductors of its
type. It is a particularly interesting case due to the relatively
short distance between nearest neighbour magnetic ions (4.7 \AA )
compared to 5.3 \AA\ for \textbf{R}Rh$_{4}$B$_{4}$ and 6.5 \AA\ for
\textbf{R}Mo$_{6}$S$_{8}$.  The proximity between the magnetic ions
prevents a clear isolation between the magnetic and superconducting
sublattices \cite{Amato}.

\yps at ambient pressure undergoes a transition into superconductivity at T$_{SC}$ = 2.3 K, and
a further transition into antiferromagnetism at T$_{N}$ = 0.22 K \cite{Aoki}.  The
antiferromagnetic order (measured using neutrons at T = 30 mK \cite{Donni}) has a
commensurate structure (FCC-type I) with an ordering wavevector
of \textbf{k} = [0, 0, 1].  Within the ab-plane the moments are ordered ferromagnetically,
but adjacent planes order antiferromagnetically.
The ordered Yb magnetic moments have magnitude m = 1.4(1) $\mu_{B}$
and are aligned along the [1, 1, 1] direction.

The superconductivity has been shown to display reentrant behaviour
 in a modest magnetic field of 500 G
(0.05 T) \cite{Aoki}, whose origin is believed to be due to the competition
 between  magnetic and superconducting
ordering fluctuations \cite{Machida2}.

\section{\label{Experiments}Experimental methods}
The ac magnetic susceptibility measurements under pressure were
undertaken in a standard BeCu Dunstan type anvil cell employing
moissanite anvils with a 1.0 mm beveled culet. The pre-indented gaskets were made of 
BeCu or phosphorus bronze.
 The pressure medium used in these cells was Daphne oil. To ensure that the samples
  were not squashed between the anvils or by the measurement coils, 
they had typical dimensions 150 $\mu$m
x 100 $\mu$m x 50 $\mu$m.

Magnetic susceptibility measurements were made using a microcoil
setup  based on the design of Alireza \cite{Alireza}. 
The diameter of the 10 turn balanced  pick-up coil was 250 - 300 $\mu$m. All wiring
was insulated from the gasket using cured Stycast 1266 with alumina.
 At room temperature, ruby was used as a
pressure gauge for all measurements.  Additionally at low
temperatures for p $\geq$ 45 kbar, the
superconducting transition temperature of lead was used as a
pressure gauge \cite{Bireckoven}. 

The four point electrical resistivity measurements were made in a piston 
cylinder cell \cite{Walker}.
In this case the pressure medium was a mixture of 50 \% n-pentane with 
50 \% iso-pentane, and lead was used to measure the pressure.

Electrical measurements were amplified using low
temperature transformers.  The signal was multiplied further
by Brookdeal EG\&G 5006 ultra-low noise pre-amplifiers.  The
measurement was taken with a lock-in amplifier. The low
temperatures were achieved using a CMR adiabatic demagnetisation
refrigerator, whose optimal base temperature was 55 mK.
The polycrystalline samples measured were cleaved from the same
ingot.

\section{\label{Results}Results}

 Ambient pressure four-point resistivity measurements
indicate that the RRR of this ingot was 12, which is
comparable to some of the better samples reported in literature.

\begin{figure}
\begin{center}
\includegraphics[width=0.45\textwidth]{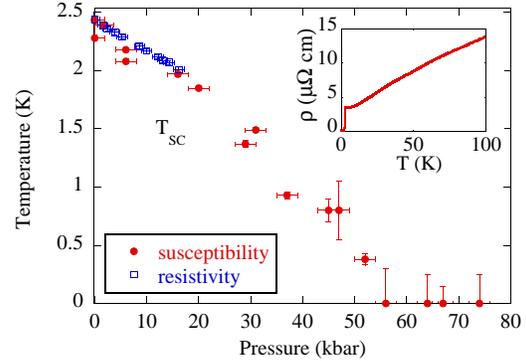}
\end{center}
\caption{\label{fig:fig1} (Color online) Main: The suppression of the superconducting transition temperature of \yps
as a function of pressure in zero applied magnetic field.  The filled 
 circles are measurements from magnetic susceptibility, 
and the open  squares are from four point resistivity.
Inset: An example of the drop observed in the ac magnetic susceptibility 
	at the onset of superconductivity in \yps \cite{Pearson}.  This data was taken at ambient pressure
 and with zero applied magnetic field.}

\end{figure}

At ambient pressure there was a clear drop in the magnetic
susceptibility signal at T = 2.4 K, which has been attributed to the
superconducting transition observed in other published work.  
An example of this signal drop is given in the inset of Fig. \ref{fig:fig1}. 
 Note that the y-axes of all of the insets to figures shown in this section display the measured
 ac magnetic susceptibility signal in arbitrary units, on a linear scale.
 
The superconducting transition was observed as a function of
pressure using both magnetic susceptibility and four point resistivity \cite{Pearson} (Fig.
\ref{fig:fig1}).
  In this figure, the superconducting transition
temperature, which was estimated as the 90 \% point of the drop, is
seen to be suppressed with pressure.  The transition appears to be
forced to below our minimum readable temperature, typically 100 mK, for pressures above 58 kbar.   Relatively large 
error bars are depicted on data points at pressures above the expected critical pressure.
These arise because the antiferromagnetic peak in the magnetic susceptibility 
data at higher temperatures (i.e. when T$_N$ $\geq$ T$_C$) means that it is difficult 
to unambiguosly rule out the possibility
of superconductivity at the lowest temperatures measured.

\begin{figure}
\begin{center}
\includegraphics[width=0.45\textwidth]{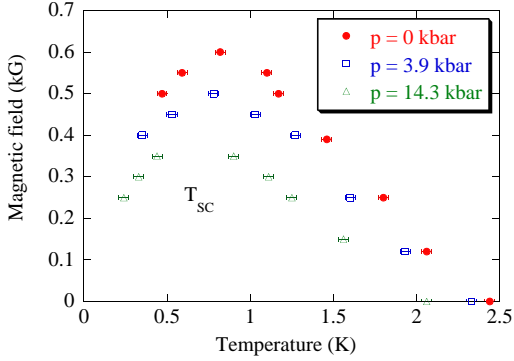}
\end{center}
\caption{ \label{fig:fig2} (Color online) The reentrant superconducting behaviour of \yps as a 
function of magnetic field for p = 0 kbar (filled circles), 3.9 kbar (open squares) and 
 14.6 kbar (open triangles),
 as measured by four point resistivity \cite{Pearson}.  The plotted data indicates the 
temperature of the 90 \% point of the drop at 
the onset or return from superconductivity.}
\end{figure}

The superconductivity displayed reentrant behaviour, and
 was observed as a function of pressure up to 14.6 kbar by four point resistivity \cite{Pearson}.
A selection of these results are shown in Fig. \ref{fig:fig2}.  
It is clearly seen that as the pressure 
is increased the field required to cause reentrance reduces.  Along with this the superconducting 
transition temperatures for both entry into superconductivity, and reentrance to the normal state, 
reduce with pressure.    

\begin{figure}
\includegraphics[width=0.45\textwidth]{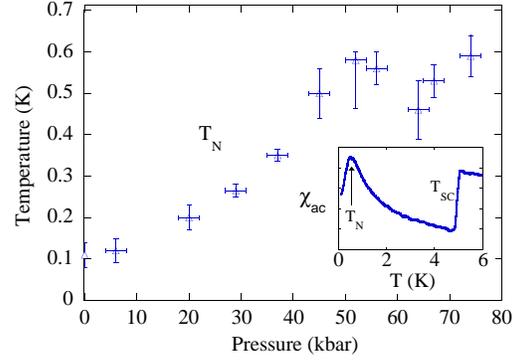}
\caption{\label{fig:fig3} (Color online) Main: The enhancement of the antiferromagnetic transition 
temperature of \yps as a function of pressure in zero applied magnetic field.  
Inset: The relative sizes of antiferromagnetic peak of \yps and the 
superconducting transition of lead at T = 4.9 K (indicating a pressure of 67 kbar).
The lead was \simi\ 12 times smaller by volume than the \yps sample. }
 
\end{figure}

    A lambda-like peak in the magnetic susceptibility has been
attributed to the onset of antiferromagnetic order, T$_{N}$, which
has been observed elsewhere \cite{Aoki}, \cite{Donni},
\cite{Kierstead}.  In this work, at ambient pressure this was
measured to occur at 0.12 K in both unpressurised moissanite anvil
cells and in standard ambient pressure balanced coils.  The value for T$_{N}$ 
at ambient pressure increases to approximately 0.25 K
 in a magentic field of 500 G.  This compares to a
temperature of 0.22 - 0.23 K in zero magnetic field 
in other work \cite{Aoki, Kierstead}.  It is
speculated here that the discrepancy made be due to differences in
the crystals measured. The motion of the antiferromagnetic
transition as a function of pressure may also be observed in Fig.
\ref{fig:fig3}.  At ambient pressure  T$_{N}$ = 0.12 K, and
rises monotonically to reach T$_{N}$ = 0.58 K at p = 74 kbar.  
Figure \ref{fig:fig3}, suggests that the
gradient of the increase in the transition temperature with pressure
shows the hint of decreasing at the higher pressures.
  Further measurements at higher
pressures would be required to confirm this idea.

    The relative magnitude of the antiferromagnetic
 peak of \yps is shown in the inset of Fig. \ref{fig:fig3}.
A clear drop in the susceptibility signal at T = 4.9 K is due to the
onset of superconductivity in the lead pressure gauge. The
temperature of this transition indicates that the measurement is at
a pressure of 64 kbar \cite{Bireckoven}.  The \yps has a volume approximately 12
times greater than the lead in this measurement.

\begin{figure}
\begin{center}
\includegraphics[width=0.45\textwidth]{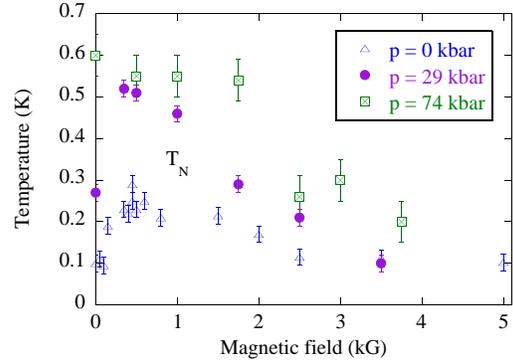}
\end{center}
\caption{ \label{fig:fig4} (Color online) The behaviour of the antiferromagnetic peak as a function of magnetic field at p = 0, 
29 and 74 kbar.  In the two lower pressure measurements (open triangles 
and squares respectively) the behaviour of 
T$_{N}$ with field shows an initial increase until B = 1 kG, then 
the transition temperature falls 
with increasing field.  The data taken at p = 74 kbar shows that 
the transition temperature 
continuously falls as magnetic field is increased. }
 
\end{figure}

Measurements of the antiferromagnetic transition were taken 
in applied magnetic fields of 0 - 5 kG.  In the lower pressures measured
it was observed that the peak in the magnetic
susceptibility goes to a higher temperature in magnetic fields of
\simi\ 500 G - 1 kG, to then be gradually suppressed.  This is demonstrated
in Fig. \ref{fig:fig4}.  This behaviour changed noticeably at
higher pressures.  Instead of having an increase in the N\'{e}el
temperature as a function of field, the N\'{e}el temperature
continuously decreased.

  \begin{figure}
\begin{center}
\includegraphics[width=0.45\textwidth]{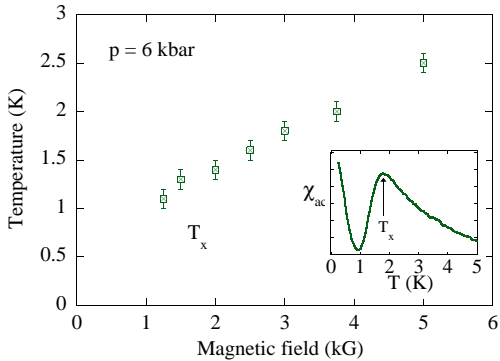}
\end{center}
\caption{ \label{fig:fig5} (Color online) Main: The motion of the unidentified transition labelled `T$_{x}$' 
	as a function of magnetic field at p = 6 kbar.
		Inset: Raw ac magnetic susceptibility data showing the shape of T$_{x}$.  The sharp 
		increase below T = 1 K is due to the residue of the 
	lambda-like peak of the antiferromagnetic transition.
	T$_{x}$ has been attributed to the peak at T = 1.8 K.}

\end{figure}

An additional feature was observed in the magnetic susceptibility
signal at all pressures above ambient pressure.  An example of this
feature is shown in the inset of Fig. \ref{fig:fig5}.  This data was
taken in an applied magnetic field of 3 kG.  To the far left of the
figure, the slope is due to the residue of the lambda-like
antiferromagnetic transition.  The new feature was the peak that is
 clearly visible at T = 1.8 K.
This peak was normally observed initially in magnetic fields \simi\
1.5 kG.  It is perceived to exist up to  B \simi\ 5 kG.  At p = 6 kbar, the
new peak is observed to occur at higher temperature as the
magnetic field is increased.   This is shown in Fig.
\ref{fig:fig5}.  The transition was observed in all of the samples
 that were pressurised in a moissanite anvil cell, 
and in a sample measured by ac susceptibility in a piston cylinder cell.
  Often the signal for this transition 
 was fairly weak, and thus difficult to track.  At all pressures measured 
between p = 6 -- 74 kbar, in B = 2.5 kG,
 the transition occurred at T$_{x}$ = 1.6 -- 1.8 K.  With magnetic field the 
transition moved to higher temperature in a similar way to the data shown at p = 6 kbar.
The type of transition that this feature may be attributed to is unknown thus far.

\section{Conclusions and Discussion}

The application of pressure to \yps causes the superconducting
transition temperature to be suppressed, such that the transition appears to occur at 
zero temperature at p $\geq$ 58 kbar.  The superconducting 
reentrance is also forced to lower temperatures and magnetic fields as pressure is increased.
 Magnetic
susceptibility measurements at pressures between 58 kbar and 74 kbar
do not indicate any further evidence of superconductivity above T = 0.2 K.

By considering the critical magnetic field associated with superconductivity, 
a simple estimate \cite{Waldram} gives the coherence length of this compound to be \simi\ 600 $\AA$.  
Further, using the residual resistivity, it is possible to estimate
the mean free path \cite{Ashcroft} to be \simi\ 200 $\AA$.  
As the mean free path is found to be significantly smaller than 
the coherence length, it is unlikely that the superconductivity is heavy fermion in nature.
This does not, however, mean that the superconductivity is necessarily 
of a standard BCS nature.  One theory presented by Machida and co-workers 
\cite{Machida2} suggests that to explain the reentrant superconducting behaviour, a more stable
 ground state may involve a Q-dependent superconducting order parameter in the antiferromagnetic state
 \cite{Machida1}.

The antiferromagnetic transition temperature of \yps is enhanced as a
function of pressure, starting at 0.12 K at ambient pressure and
rising to 0.58 K at p = 74 kbar.  This fact is interesting because 
 neutron data  at ambient pressure suggests that at T = 0.6 K a small magnetic moment 
is observed, which increases significantly as the temperature 
reduces to $\sim$ 0.1 K, at which point
the moment doesn't increase much further \cite{Bernhoeft}. 

\begin{figure}
\begin{center}
\includegraphics[width=0.45\textwidth]{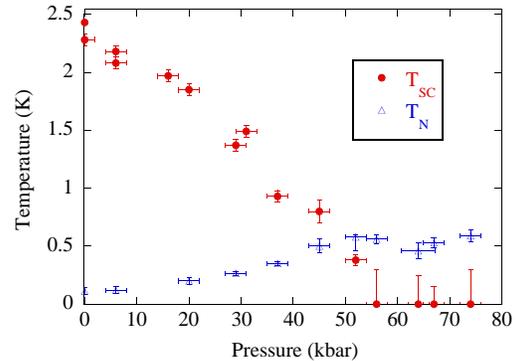}
\end{center}
\caption{\label{fig:fig6} (Color online) The phase diagram of \yps in zero magnetic field.  
The superconducting transition temperature
    is shown as T$_{SC}$, and the antiferromagnetic transition is shown by T$_{N}$.}
   
\end{figure}

The phase diagram of \yps in zero magnetic field is shown in Fig. 
\ref{fig:fig6}.  It is interesting that superconductivity survives in 
the presence of strong magnetic fluctuations in the temperature range T$_C$ $ \geq $
T $\geq$ T$_N$.  In this temperature range the long range 
magnetic order has not fully set in, but the magnetic fluctuations will be significant.
This is thought to be because the Yb sublattice, responsible for magnetism, 
is decoupled from the Pd sublattice, which carries the superconducting electrons.  This 
idea is corroborated by band structure calculations undertaken by Nevidomskyy \cite{Andriy}.

To address the question of the lack of heavy fermion superconductivity in 
Yb based compounds, one
may consider the magnitude of T$_N$ in these materials.  The 
data shown here for \yps suggests that 
T$_N$ reaches a peak at around T = 0.6 K or slightly higher.  
Compare this with the situation
in YbRh$_2$Si$_2$, which displays a maximum magnetic 
ordering temperature of around T = 1 K at p = 40 kbar \cite{Knebel}.
The analogous Ce based compound is CeRh$_2$Si$_2$.  
This compound has an antiferromagnetic transition at p = 0 kbar
at T$_N$ = 36 K \cite{Nishigoria}.  As the strength 
of the relevant RKKY magnetic coupling constant,
J$_{RKKY}$, is related to the maximum value of 
T$_N$ (as in the Doniach phase diagram \cite{Doniach}), the measurements of 
\textbf{R}Rh$_2$Si$_2$ (\textbf{R} = Yb, Ce) suggest 
that the RKKY coupling constant in
Yb systems may be lower than in Ce systems.  This 
is consistent with Yb systems 
having nearly local moment magnetism. 
This is supported by the data in this work, 
showing the possibility of another Yb compound with a low T$_N$ maximum  
value with pressure.  As heavy fermion superconductivity
relies on a larger value of J$_{RKKY}$ \cite{Doniach2}, 
the lower J$_{RKKY}$ values in Yb compounds may account for the absence 
of superconductivity in the temperature and compound purity ranges currently possible experimentally.

Evidence in the ac magnetic susceptibility data 
for a possible new transition in \yps has been found at p $\geq$ 6 kbar
and in magnetic fields between \simi\ 1.5 kG and 5 kG.  The
nature of this transition is not known.

\begin{acknowledgments}
The authors would like to
thank S. Brown for his technical assistance, along with R. P. Smith, S. E. Rowley and I. R. Walker for useful discussions.
  This work was funded by the EPSRC and Trinity College, Cambridge.  
\end{acknowledgments}
\bibliography{bibliography}

\begin{thebibliography}{23}
\expandafter\ifx\csname natexlab\endcsname\relax\def\natexlab#1{#1}\fi
\expandafter\ifx\csname bibnamefont\endcsname\relax
  \def\bibnamefont#1{#1}\fi
\expandafter\ifx\csname bibfnamefont\endcsname\relax
  \def\bibfnamefont#1{#1}\fi
\expandafter\ifx\csname citenamefont\endcsname\relax
  \def\citenamefont#1{#1}\fi
\expandafter\ifx\csname url\endcsname\relax
  \def\url#1{\texttt{#1}}\fi
\expandafter\ifx\csname urlprefix\endcsname\relax\def\urlprefix{URL }\fi
\providecommand{\bibinfo}[2]{#2}
\providecommand{\eprint}[2][]{\url{#2}}

\bibitem[{\citenamefont{Thalmeier et~al.}(2004)\citenamefont{Thalmeier,
  Zwicknagl, Stockert, Sparn, and Steglich}}]{Thalmeier}
\bibinfo{author}{\bibfnamefont{P.}~\bibnamefont{Thalmeier}},
  \bibinfo{author}{\bibfnamefont{G.}~\bibnamefont{Zwicknagl}},
  \bibinfo{author}{\bibfnamefont{O.}~\bibnamefont{Stockert}},
  \bibinfo{author}{\bibfnamefont{G.}~\bibnamefont{Sparn}}, \bibnamefont{and}
  \bibinfo{author}{\bibfnamefont{F.}~\bibnamefont{Steglich}},
  \emph{\bibinfo{title}{Superconductivity in heavy fermion compounds}}
  (\bibinfo{publisher}{Springer, Berlin}, \bibinfo{year}{2004}),
  \bibinfo{edition}{1st} ed.

\bibitem[{\citenamefont{Onuki et~al.}(2004)\citenamefont{Onuki, Settai,
  Sugiyama, Takeuchi, Kobayashi, Haga, and Yamamoto}}]{Onuki}
\bibinfo{author}{\bibfnamefont{Y.}~\bibnamefont{Onuki}},
  \bibinfo{author}{\bibfnamefont{R.}~\bibnamefont{Settai}},
  \bibinfo{author}{\bibfnamefont{K.}~\bibnamefont{Sugiyama}},
  \bibinfo{author}{\bibfnamefont{T.}~\bibnamefont{Takeuchi}},
  \bibinfo{author}{\bibfnamefont{T.~C.} \bibnamefont{Kobayashi}},
  \bibinfo{author}{\bibfnamefont{Y.}~\bibnamefont{Haga}}, \bibnamefont{and}
  \bibinfo{author}{\bibfnamefont{E.}~\bibnamefont{Yamamoto}},
  \bibinfo{journal}{J. Phys. Soc. Jpn.} \textbf{\bibinfo{volume}{73}},
  \bibinfo{pages}{769} (\bibinfo{year}{2004}).

\bibitem[{\citenamefont{Thompson et~al.}(2001)\citenamefont{Thompson,
  Movshovich, Fisk, Bouquet, Curro, Fisher, Hammel, Hegger, Hundley, Jaime
  et~al.}}]{Thompson}
\bibinfo{author}{\bibfnamefont{J.~D.} \bibnamefont{Thompson}},
  \bibinfo{author}{\bibfnamefont{R.}~\bibnamefont{Movshovich}},
  \bibinfo{author}{\bibfnamefont{Z.}~\bibnamefont{Fisk}},
  \bibinfo{author}{\bibfnamefont{F.}~\bibnamefont{Bouquet}},
  \bibinfo{author}{\bibfnamefont{N.~J.} \bibnamefont{Curro}},
  \bibinfo{author}{\bibfnamefont{R.}~\bibnamefont{Fisher}},
  \bibinfo{author}{\bibfnamefont{P.~C.} \bibnamefont{Hammel}},
  \bibinfo{author}{\bibfnamefont{H.}~\bibnamefont{Hegger}},
  \bibinfo{author}{\bibfnamefont{M.~F.} \bibnamefont{Hundley}},
  \bibinfo{author}{\bibfnamefont{M.}~\bibnamefont{Jaime}},
  \bibnamefont{et~al.}, \bibinfo{journal}{J. Magn. Magn. Mater.}
  \textbf{\bibinfo{volume}{226}}, \bibinfo{pages}{5} (\bibinfo{year}{2001}).

\bibitem[{\citenamefont{Mathur et~al.}(1998)\citenamefont{Mathur, Grosche,
  Julian, Walker, Freye, Haselwimmer, and Lonzarich}}]{Mathur}
\bibinfo{author}{\bibfnamefont{N.~D.} \bibnamefont{Mathur}},
  \bibinfo{author}{\bibfnamefont{F.~M.} \bibnamefont{Grosche}},
  \bibinfo{author}{\bibfnamefont{S.~R.} \bibnamefont{Julian}},
  \bibinfo{author}{\bibfnamefont{I.~R.} \bibnamefont{Walker}},
  \bibinfo{author}{\bibfnamefont{D.~M.} \bibnamefont{Freye}},
  \bibinfo{author}{\bibfnamefont{R.~K.~W.} \bibnamefont{Haselwimmer}},
  \bibnamefont{and} \bibinfo{author}{\bibfnamefont{G.~G.}
  \bibnamefont{Lonzarich}}, \bibinfo{journal}{Nature}
  \textbf{\bibinfo{volume}{394}}, \bibinfo{pages}{39} (\bibinfo{year}{1998}).

\bibitem[{\citenamefont{Kierstead et~al.}(1985)\citenamefont{Kierstead, Dunlap,
  Malik, Umarli, and Shenoy}}]{Kierstead}
\bibinfo{author}{\bibfnamefont{H.~A.} \bibnamefont{Kierstead}},
  \bibinfo{author}{\bibfnamefont{B.~D.} \bibnamefont{Dunlap}},
  \bibinfo{author}{\bibfnamefont{S.~K.} \bibnamefont{Malik}},
  \bibinfo{author}{\bibfnamefont{A.~M.} \bibnamefont{Umarli}},
  \bibnamefont{and} \bibinfo{author}{\bibfnamefont{G.~K.}
  \bibnamefont{Shenoy}}, \bibinfo{journal}{Phys. Rev. B}
  \textbf{\bibinfo{volume}{32}}, \bibinfo{pages}{135} (\bibinfo{year}{1985}).

\bibitem[{\citenamefont{Machida
  et~al.}(1980{\natexlab{a}})\citenamefont{Machida, Nokura, and
  Matsubara}}]{Machida1}
\bibinfo{author}{\bibfnamefont{K.}~\bibnamefont{Machida}},
  \bibinfo{author}{\bibfnamefont{K.}~\bibnamefont{Nokura}}, \bibnamefont{and}
  \bibinfo{author}{\bibfnamefont{T.}~\bibnamefont{Matsubara}},
  \bibinfo{journal}{Phys. Rev. Lett.} \textbf{\bibinfo{volume}{44}},
  \bibinfo{pages}{821} (\bibinfo{year}{1980}{\natexlab{a}}).

\bibitem[{\citenamefont{Stanley et~al.}(1987)\citenamefont{Stanley, Lynn,
  Shelton, and Klavins}}]{Stanley}
\bibinfo{author}{\bibfnamefont{H.~B.} \bibnamefont{Stanley}},
  \bibinfo{author}{\bibfnamefont{J.~W.} \bibnamefont{Lynn}},
  \bibinfo{author}{\bibfnamefont{R.~N.} \bibnamefont{Shelton}},
  \bibnamefont{and} \bibinfo{author}{\bibfnamefont{P.}~\bibnamefont{Klavins}},
  \bibinfo{journal}{J. Appl. Phys.} \textbf{\bibinfo{volume}{61}},
  \bibinfo{pages}{3371} (\bibinfo{year}{1987}).

\bibitem[{\citenamefont{Aoki et~al.}(2000)\citenamefont{Aoki, Sato, Sugawara,
  and Sato}}]{Aoki}
\bibinfo{author}{\bibfnamefont{Y.}~\bibnamefont{Aoki}},
  \bibinfo{author}{\bibfnamefont{H.~R.} \bibnamefont{Sato}},
  \bibinfo{author}{\bibfnamefont{H.}~\bibnamefont{Sugawara}}, \bibnamefont{and}
  \bibinfo{author}{\bibfnamefont{H.}~\bibnamefont{Sato}},
  \bibinfo{journal}{Physica C} \textbf{\bibinfo{volume}{333}},
  \bibinfo{pages}{187} (\bibinfo{year}{2000}).

\bibitem[{\citenamefont{Amato et~al.}(2003)\citenamefont{Amato, Roessli,
  Fischer, Bernhoeft, Stunault, Baines, D\"{o}nni, and Sugawara}}]{Amato}
\bibinfo{author}{\bibfnamefont{A.}~\bibnamefont{Amato}},
  \bibinfo{author}{\bibfnamefont{B.}~\bibnamefont{Roessli}},
  \bibinfo{author}{\bibfnamefont{P.}~\bibnamefont{Fischer}},
  \bibinfo{author}{\bibfnamefont{N.}~\bibnamefont{Bernhoeft}},
  \bibinfo{author}{\bibfnamefont{A.}~\bibnamefont{Stunault}},
  \bibinfo{author}{\bibfnamefont{C.}~\bibnamefont{Baines}},
  \bibinfo{author}{\bibfnamefont{A.}~\bibnamefont{D\"{o}nni}},
  \bibnamefont{and} \bibinfo{author}{\bibfnamefont{H.}~\bibnamefont{Sugawara}},
  \bibinfo{journal}{Physica B} \textbf{\bibinfo{volume}{326}},
  \bibinfo{pages}{369} (\bibinfo{year}{2003}).

\bibitem[{\citenamefont{D\"{o}nni et~al.}(1999)\citenamefont{D\"{o}nni,
  Fischer, Fauth, Convert, Aoki, Sugawara, and Sato}}]{Donni}
\bibinfo{author}{\bibfnamefont{A.}~\bibnamefont{D\"{o}nni}},
  \bibinfo{author}{\bibfnamefont{P.}~\bibnamefont{Fischer}},
  \bibinfo{author}{\bibfnamefont{F.}~\bibnamefont{Fauth}},
  \bibinfo{author}{\bibfnamefont{P.}~\bibnamefont{Convert}},
  \bibinfo{author}{\bibfnamefont{Y.}~\bibnamefont{Aoki}},
  \bibinfo{author}{\bibfnamefont{H.}~\bibnamefont{Sugawara}}, \bibnamefont{and}
  \bibinfo{author}{\bibfnamefont{H.}~\bibnamefont{Sato}},
  \bibinfo{journal}{Physica B} \textbf{\bibinfo{volume}{259-261}},
  \bibinfo{pages}{705} (\bibinfo{year}{1999}).

\bibitem[{\citenamefont{Machida
  et~al.}(1980{\natexlab{b}})\citenamefont{Machida, Nokura, and
  Matsubara}}]{Machida2}
\bibinfo{author}{\bibfnamefont{K.}~\bibnamefont{Machida}},
  \bibinfo{author}{\bibfnamefont{K.}~\bibnamefont{Nokura}}, \bibnamefont{and}
  \bibinfo{author}{\bibfnamefont{T.}~\bibnamefont{Matsubara}},
  \bibinfo{journal}{Phys. Rev. B} \textbf{\bibinfo{volume}{22}},
  \bibinfo{pages}{2307} (\bibinfo{year}{1980}{\natexlab{b}}).

\bibitem[{\citenamefont{Alireza and Julian}(2003)}]{Alireza}
\bibinfo{author}{\bibfnamefont{P.~L.} \bibnamefont{Alireza}} \bibnamefont{and}
  \bibinfo{author}{\bibfnamefont{S.~R.} \bibnamefont{Julian}},
  \bibinfo{journal}{Rev. Sci. Instrum.} \textbf{\bibinfo{volume}{74}},
  \bibinfo{pages}{4728} (\bibinfo{year}{2003}).

\bibitem[{\citenamefont{Bireckoven and Wittig}(1988)}]{Bireckoven}
\bibinfo{author}{\bibfnamefont{B.}~\bibnamefont{Bireckoven}} \bibnamefont{and}
  \bibinfo{author}{\bibfnamefont{J.}~\bibnamefont{Wittig}},
  \bibinfo{journal}{J. Phys. E} \textbf{\bibinfo{volume}{21}},
  \bibinfo{pages}{841} (\bibinfo{year}{1988}).

\bibitem[{\citenamefont{Walker}(1999)}]{Walker}
\bibinfo{author}{\bibfnamefont{I.~R.} \bibnamefont{Walker}},
  \bibinfo{journal}{Rev. Sci. Instrum.} \textbf{\bibinfo{volume}{70}},
  \bibinfo{pages}{3402} (\bibinfo{year}{1999}).

\bibitem[{\citenamefont{Pearson}(2005)}]{Pearson}
\bibinfo{author}{\bibfnamefont{E.~E.} \bibnamefont{Pearson}}, Ph.D. thesis,
  \bibinfo{school}{University of Cambridge} (\bibinfo{year}{2005}).

\bibitem[{\citenamefont{Waldram}(1996)}]{Waldram}
\bibinfo{author}{\bibfnamefont{J.~R.} \bibnamefont{Waldram}},
  \emph{\bibinfo{title}{Superconductivity of Metals and Cuprates}}
  (\bibinfo{publisher}{IOP Publishing Ltd.}, \bibinfo{year}{1996}),
  \bibinfo{edition}{1st} ed.

\bibitem[{\citenamefont{Ashcroft and Mermin}(1976)}]{Ashcroft}
\bibinfo{author}{\bibfnamefont{N.~W.} \bibnamefont{Ashcroft}} \bibnamefont{and}
  \bibinfo{author}{\bibfnamefont{N.~D.} \bibnamefont{Mermin}},
  \emph{\bibinfo{title}{Solid State Physics}}
  (\bibinfo{publisher}{Brooks/Cole}, \bibinfo{year}{1976}),
  \bibinfo{edition}{1st} ed.

\bibitem[{\citenamefont{Bernhoeft}()}]{Bernhoeft}
\bibinfo{author}{\bibfnamefont{N.}~\bibnamefont{Bernhoeft}},
  \bibinfo{howpublished}{Private Communication}.

\bibitem[{\citenamefont{Nevidomskyy}()}]{Andriy}
\bibinfo{author}{\bibfnamefont{A.}~\bibnamefont{Nevidomskyy}},
  \bibinfo{howpublished}{Private Communication}.

\bibitem[{\citenamefont{Knebel et~al.}(2005)\citenamefont{Knebel, Glazkov,
  Pourret, Niklowitz, Lapertot, Salce, and Flouquet}}]{Knebel}
\bibinfo{author}{\bibfnamefont{G.}~\bibnamefont{Knebel}},
  \bibinfo{author}{\bibfnamefont{V.}~\bibnamefont{Glazkov}},
  \bibinfo{author}{\bibfnamefont{A.}~\bibnamefont{Pourret}},
  \bibinfo{author}{\bibfnamefont{P.~G.} \bibnamefont{Niklowitz}},
  \bibinfo{author}{\bibfnamefont{G.}~\bibnamefont{Lapertot}},
  \bibinfo{author}{\bibfnamefont{B.}~\bibnamefont{Salce}}, \bibnamefont{and}
  \bibinfo{author}{\bibfnamefont{J.}~\bibnamefont{Flouquet}},
  \bibinfo{journal}{Physica B} \textbf{\bibinfo{volume}{359-361}},
  \bibinfo{pages}{20} (\bibinfo{year}{2005}).

\bibitem[{\citenamefont{Nishigori et~al.}(2005)\citenamefont{Nishigori,
  Miyamoto, Ikeda, and Ito}}]{Nishigoria}
\bibinfo{author}{\bibfnamefont{S.}~\bibnamefont{Nishigori}},
  \bibinfo{author}{\bibfnamefont{N.}~\bibnamefont{Miyamoto}},
  \bibinfo{author}{\bibfnamefont{T.}~\bibnamefont{Ikeda}}, \bibnamefont{and}
  \bibinfo{author}{\bibfnamefont{T.}~\bibnamefont{Ito}},
  \bibinfo{journal}{Physica B} \textbf{\bibinfo{volume}{359-361}},
  \bibinfo{pages}{172} (\bibinfo{year}{2005}).

\bibitem[{\citenamefont{Doniach}(1977)}]{Doniach}
\bibinfo{author}{\bibfnamefont{S.}~\bibnamefont{Doniach}},
  \bibinfo{journal}{Physica B} \textbf{\bibinfo{volume}{91}},
  \bibinfo{pages}{231} (\bibinfo{year}{1977}).

\bibitem[{\citenamefont{Doniach and Sondheimer}(1998)}]{Doniach2}
\bibinfo{author}{\bibfnamefont{S.}~\bibnamefont{Doniach}} \bibnamefont{and}
  \bibinfo{author}{\bibfnamefont{E.~H.} \bibnamefont{Sondheimer}},
  \emph{\bibinfo{title}{Green's functions for solid state physicists}}
  (\bibinfo{publisher}{Imperial College Press}, \bibinfo{year}{1998}).

\end{thebibliography}

\end{document}